\documentstyle[12pt]{article}
\begin{document}
\vskip .7cm
\begin{center}
{\bf { NEW LOCAL SYMMETRY FOR QED IN TWO DIMENSIONS}}

\vskip 2cm

 R. P. MALIK
\footnote{ E-mail: malik@boson.bose.res.in  }\\
{\it S. N. Bose National Centre for Basic Sciences,} \\
{\it Block-JD, Sector-III, Salt Lake, Calcutta-700 098, India} \\

\vskip 2.5cm

\end{center}
{\bf Abstract:}               
A new local, covariant and nilpotent
symmetry is shown to exist for the interacting BRST invariant $U(1)$ 
gauge theory in two dimensions of space-time. 
Under this new symmetry, it is the gauge-fixing term that remains invariant
and the corresponding transformations on the Dirac fields turn out to be the
analogue of chiral transformations.  The extended BRST algebra is derived 
for the generators of all the underlying symmetries, present in the theory. 
This algebra turns out to be the analogue of the algebra obeyed by the 
de Rham cohomology operators of differential geometry.
Possible interpretations and implications of this symmetry are pointed 
out in the context of BRST cohomology and Hodge decomposition theorem. 

\baselineskip=16pt

\vskip 1cm

\newpage

\noindent
The importance of symmetries in  physics has gone beyond 
the mere requirement of an aesthetic appeal to the sophistication of 
explaining some of the landmark experiments. One such important
symmetry in theoretical physics is the gauge symmetry which 
has turned out to provide the corner-stone for the modern
developments in the ideas of unification. The main characteristic feature of
theories, based on these symmetries, is the fact that they are endowed 
with the first-class constraints (in the language of Dirac) [1-3]. 
A cardinal example of such a class of theories
is the quantum electrodynamics (QED) which represents a dynamically closed
system of photon ($U(1)$ gauge field $A_{\mu}$) and electrically charged
particles (e.g. electrons and positrons which are described by the
Dirac fields $\psi$ and $\bar \psi$). For the covariant canonical quantization
of these gauge theories, the Becchi-Rouet-Stora-Tyutin (BRST) 
symmetry [4,5] turns out to be quite handy. In this formalism,  unitarity and 
gauge invariance are respected together at any arbitrary order of
perturbation theory. The purpose of the present letter is to shed some light 
on the existence of some local, covariant and continuous symmetries
which have not been explored hitherto in the context of two dimensional (2D) 
QED and to establish that this {\it interacting} theory provides 
{\it a physically tractable field theoretical model for the Hodge theory}. The 
generators for the underlying symmetries of the theory obey an algebra that is 
reminiscent of the algebra of de Rham cohomology operators of differential 
geometry defined on a compact manifold.

Let us begin with a 
$D$-dimensional BRST invariant Lagrangian density (${\cal L}_{B}$) for
the interacting $U(1)$ gauge theory in the Feynman gauge:
$$
\begin{array}{lcl}
{\cal L}_{B} = - \frac{1}{4} F^{\mu\nu} F_{\mu\nu}
+ \bar \psi (i \gamma^{\mu} \partial_{\mu} - m) \psi
- e \bar \psi \gamma^{\mu} A_{\mu} \psi
+ B (\partial \cdot A) + \frac{1}{2} B^2 - i \partial_{\mu} \bar C
\partial^{\mu} C,
\end{array}\eqno (1)
$$
where $F_{\mu\nu} = \partial_{\mu} A_{\nu} - \partial_{\nu} A_{\mu}$
is the field strength tensor, $B$ is the Nakanishi-Lautrup auxiliary field,
$(\bar C)C$ are the Faddeev-Popov (anti)ghost fields ($ \bar C^2 
= C^2 = 0$) and
indices $\mu,\nu = 0, 1, 2,......D-1$ represent the flat Minkowski space-time
directions. It can be checked that the above Lagrangian density remains
quasi-invariant ($\delta_{B} {\cal L}_{B} = \eta \partial_{\mu}
[B \partial^{\mu} C]$) under the following off-shell nilpotent
($ \delta_{B}^2 = 0$) BRST transformations:    
$$
\begin{array}{lcl}
\delta_{B} A_{\mu} &=& \eta\; \partial_{\mu}\; C, \;\;\qquad \;\;
\delta_{B} C = 0, \; \;\qquad \; \;\delta_{B} \bar C 
= i \;\eta\; B, \;\;\qquad \;\;
\delta_{B} B = 0, \nonumber\\
\delta_{B} \psi &=& - i \eta e C \psi, \quad\;
\delta_{B} \bar \psi = i \eta e C \bar \psi, \quad \;
\delta_{B} (\partial \cdot A) = \eta \Box C, \quad \;
\delta_{B} F_{\mu\nu} = 0,
\end{array}\eqno (2)
$$
where $\eta$ is an anticommuting ($ \eta C = - C \eta, \eta \bar C =
- \bar C \eta, \eta \psi = - \psi \eta, \eta \bar \psi = - \bar \psi \eta$)
space-time independent transformation parameter. Using Noether theorem, it
is straightforward to check that the generator for the above transformations
is the nilpotent ($Q_{B}^2 = 0$) BRST charge
$$
\begin{array}{lcl}
Q_{B} &=& {\displaystyle \int} d^{(D-1)} x\;
\bigl [\; \partial_{i} F^{0i} C + B \dot C - e \bar \psi \gamma_{0} C \psi
\;\bigr ], \nonumber\\
&\equiv& {\displaystyle \int} d^{(D-1)} x \;\bigl [ \;B \dot C - \dot B C
\;\bigr ],
\end{array} \eqno(3)
$$
where the latter expression for $Q_{B}$ has been obtained by exploiting
the equation of motion ($ \partial_{\mu} F^{\mu\nu} - \partial^{\nu} B
= e \bar \psi \gamma^{\nu} \psi $). The global scale invariance of (1)
under $ C \rightarrow e^{- \lambda} C, \bar C \rightarrow e^{\lambda}
\bar C, A_{\mu} \rightarrow A_{\mu}, B \rightarrow B $ (where $\lambda$
is a global parameter), leads to the derivation of a conserved ghost charge
($Q_{g}$)
$$
\begin{array}{lcl}
Q_{g} 
= - i {\displaystyle \int} d^{(D-1)} x \;\bigl [\; C \dot {\bar C} 
+ \bar C \dot  C
\; \bigr ].
\end{array} \eqno(4)
$$
Together, these conserved charges satisfy the following algebra
$$
\begin{array}{lcl}
Q_{B}^2 &=& \frac{1}{2} \{ Q_{B}, Q_{B} \} = 0, \;\;\qquad\;\;
i [ Q_{g}, Q_{B} ] = + Q_{B}, \nonumber\\
Q_{AB}^2 &=& \frac{1}{2} \{ Q_{AB}, Q_{AB} \} = 0, \qquad
i [ Q_{g}, Q_{AB} ] = - Q_{AB},
\end{array} \eqno(5)
$$
where $Q_{AB}$ is the anti-BRST charge which can be readily obtained from (3)
by the replacement :$ C \rightarrow i \bar C$ [6-8]. Note that
$ C \rightarrow \pm i \bar C, \bar C \rightarrow \pm i C$ is the
symmetry of the ghost action ($ I_{F.P.} = - i \int d^{D} x\;
\partial_{\mu} \bar C \partial^{\mu} C$)
in any arbitrary dimension of space-time.

In the past few years, many authors [9-12] have attempted to explore the 
possibilities of obtaining new BRST-type symmetries for the Lagrangian 
density (1) of QED in the hope of establishing a deeper connection
with the mathematical aspects of BRST cohomology in any arbitrary dimension
of space-time (see, e.g., Refs. [9,12]). However, the symmetry
transformations turn out to be nonlocal and noncovariant. In the
relativistic covariant formulation of these symmetries [13] the manifest
nilpotency is lost and it is restored only under certain specific
restrictions on the parameters of the theory. The central theme of our 
present short note is to show that in two dimensions of space-time there 
exists a local, continuous, covariant and nilpotent BRST-type symmetry
under which the gauge-fixing term of the Lagrangian density (1)
remains invariant and corresponding transformations on Dirac fields
turn out to be the analogue of chiral transformations. 
This symmetry transformation is not the generalization
of the $D$-dimensional symmetries [9-12] explored by others 
to two dimensions of space-time. {\it Rather, it is a new symmetry in its own 
right}. Contrary to the above symmetry transformation, it is the 
Abelian field strength tensor (two-form)
that remains invariant under the usual BRST 
symmetry and the Dirac fields transform as the analogue
of gauge transformations (see, e.g., eqn.(2)). We christen
the new symmetry as the dual-BRST symmetry because the gauge-fixing term is 
Hodge dual to the field strength tensor (two-form) of Abelian $U(1)$ 
gauge theory in any arbitrary dimension of space-time
\footnote{The vector potential $A_{\mu}$ of the $U(1)$ gauge theory
is defined through one-form $A= A_{\mu} dx^{\mu}$. The gauge-fixing
term $ \partial \cdot A = \delta A $ is Hodge dual to the two-form $F 
= d A $ where $ \delta = \pm * d * $ is the adjoint(dual) 
exterior derivative and $d$ is the exterior derivative (see, e.g.,
Ref. [15]).} [14,15]. This duality is also reflected at the level of 
transformations for the Dirac fields where the analogues of gauge- 
and chiral transformations are dual to each-other.

In two ($ 1 + 1$) dimensions of space-time, there exists only one component
(i.e. electric field $E = F_{01}$) of the field 
strength tensor $F_{\mu\nu}$.
Thus, the analogue of the BRST invariant Lagrangian density (1) is:
$$
\begin{array}{lcl}
{\cal L}_{B} =  \frac{1}{2} \; E^2
+ \bar \psi (i \gamma^{\mu} \partial_{\mu} - m) \psi
- e \bar \psi \gamma^{\mu} A_{\mu} \psi
+ B (\partial \cdot A) + \frac{1}{2} B^2 - i \partial_{\mu} \bar C
\partial^{\mu} C,
\end{array}\eqno (6a)
$$
which can be recast as:
$$
\begin{array}{lcl}
{\cal L_{B}} =  {\cal B} E - \frac{1}{2} \; {\cal B}^2
+ \bar \psi (i \gamma^{\mu} \partial_{\mu} - m) \psi
- e \bar \psi \gamma^{\mu} A_{\mu} \psi
+ B (\partial \cdot A) + \frac{1}{2} B^2 - i \partial_{\mu} \bar C
\partial^{\mu} C,
\end{array}\eqno (6b)
$$
by introducing another auxiliary field ${\cal B}$. It can be checked that
under the following off-shell nilpotent ($\delta_{D}^2 = 0$) dual-BRST 
transformations
$$
\begin{array}{lcl}
\delta_{D} A_{\mu} &=& -\eta \varepsilon_{\mu\nu}
\partial^{\nu} \bar C, \quad\;
\delta_{D} \bar C = 0, \quad \;\delta_{D}  C = - i \eta {\cal B}, \quad\;
\delta_{D} {\cal B} = 0, \quad \;\delta_{D} B = 0, \nonumber\\
\delta_{D} \psi &=& - i \eta e \bar C \gamma_{5} \psi, \quad\;
\delta_{D} \bar \psi = i \eta e \bar C \gamma_{5} \bar \psi, \quad\;
\delta_{D} (\partial \cdot A) = 0, \quad\;
\delta_{D} E = \eta \Box {\bar C}, 
\end{array}\eqno (7)
$$
the Lagrangian density (6b) (with $m = 0$) transforms as:
$\delta_{D} {\cal L_{B}} = \eta \partial_{\mu} ({\cal B} \partial^{\mu}
\bar C)$
\footnote{ We adopt the notations in which the flat 2D Minkowski metric
$\eta_{\mu\nu} = $ diag$\; (+1, -1), \gamma^{0} = \sigma_{2}, 
\gamma^{1} = i \sigma_{1}, \gamma_{5} = \gamma^{0} \gamma^{1} = \sigma_{3},
\{ \gamma^{\mu}, \gamma^{\nu} \} = 2 \eta^{\mu\nu}, \gamma_{\mu} \gamma_{5}
= \varepsilon_{\mu\nu} \gamma^{\nu}, \varepsilon_{01} = \varepsilon^{10}
= + 1, F_{01} = \partial_{0} A_{1} - \partial_{1} A_{0} 
= E = - \varepsilon^{\mu\nu} \partial_{\mu} A_{\nu} = F^{10},
\Box = \eta^{\mu\nu} \partial_{\mu} \partial_{\nu} = (\partial_{0})^2
- (\partial_{1})^2 $ and  here $\sigma's$ are the usual
$ 2 \times 2$ Pauli matrices.}. Exploiting
the Noether theorem, it can be checked that the above transformations
are generated by
$$
\begin{array}{lcl}
Q_{D} &=& {\displaystyle \int} d x\;
\bigl [\; {\cal B} \dot {\bar C} + e \bar \psi \gamma_{1} \bar C \psi
- (\partial_{1} B) \bar C\;
\bigr ], \nonumber\\
&\equiv& {\displaystyle \int} d x \;\bigl [ \;{\cal B} \dot {\bar C} 
- \dot {\cal B} \bar C
\;\bigr ],
\end{array} \eqno(8)
$$
where the latter expression for $Q_{D}$ has been obtained by using
the equation of motion ($ \varepsilon^{\mu\nu} \partial_{\nu}
{\cal B} + \partial^{\mu} B = - e \bar \psi \gamma^{\mu} \psi$) for the
photon field, present in the Lagrangian density (6b). 
Using the following BRST quantization conditions (with $\hbar = c = 1$):
$$
\begin{array}{lcl}
&& [ A_{0} (x,t), B (y, t) ] = i \delta (x - y), \nonumber\\
&& [ A_{1} (x, t), {\cal B} (y, t) ] = i \delta (x - y), \nonumber\\
&& \{ \psi (x, t), \psi^{\dagger} (y, t) \} = - \delta (x - y), \nonumber\\
&& \{ C (x ,t), \dot {\bar C} (y, t) \} = \delta (x - y), \nonumber\\
&& \{ \bar C (x, t), \dot C(y, t) \} = - \delta (x -y), 
\end{array} \eqno(9)
$$
(and rest of the (anti)commutators are zero), it can be seen
that $Q_{D}$ is
indeed the generator for the transformations (7) if we exploit the
following relationship 
$$
\begin{array}{lcl}
\delta_{D} \Phi = - i \eta [ \Phi, Q_{D} ]_{\pm},
\end{array}\eqno(10)
$$
where $[\;,\;]_{\pm}$ stands for (anti)commutator for the generic field
$\Phi$ being (fermionic)bosonic in nature. It is straightforward to check
that $ Q_{D}^2 = \frac{1}{2} \{ Q_{D}, Q_{D} \} = 0$ due to (9). A simpler
way to see this fact is: $ \delta_{D} Q_{D} = - i \eta \{ Q_{D}, Q_{D} \} = 0$
by exploiting (7) and (8).

It is very natural to expect that the anticommutator of these two
transformations ($\{ \delta_{B}, \delta_{D} \} = \delta_{W}$) would
also be the symmetry transformation ($\delta_{W}$) for the Lagrangian
density (6b) (with $ m = 0$). This is indeed the case as can be seen
that under the following bosonic ($ \kappa = - i \eta \eta^{\prime}$)
transformations corresponding to $\delta_{W}$
$$
\begin{array}{lcl}
\delta_{W} A_{\mu} &=& \kappa (\partial_{\mu} {\cal B}
+ \varepsilon_{\mu\nu} \partial^{\nu} B), \qquad\;
\delta_{W} {\cal B} = 0, \qquad \;\delta_{W} B = 0,
\nonumber\\
\delta_{W} (\partial \cdot A) &=& \kappa \Box {\cal B}, \;\qquad
\delta_{W} E = - \kappa \Box B, \;\quad 
\delta_{W} C = 0, \;\quad \delta_{W} \bar C = 0,
\nonumber\\
\delta_{W} \psi &=& \kappa i e (\gamma_{5} B - {\cal B}) \psi, \;\qquad\;
\delta_{W} \bar \psi = - \kappa i e (\gamma_{5} B - {\cal B}) \bar \psi,
\end{array}\eqno(11)
$$
the Lagrangian density (6b) (with $ m = 0$) transforms as:
$ \delta_{W} {\cal L_{B}} = \kappa \partial_{\mu} [ B \partial^{\mu}
{\cal B} - {\cal B} \partial^{\mu} B ]$. Here $\eta$ and
$\eta^{\prime}$ are the transformation parameters corresponding to
the transformations $\delta_{B}$ and $\delta_{D}$ respectively. 
The generator for the above transformations is:
$$
\begin{array}{lcl}
W &=& {\displaystyle \int} dx \;
\bigl [\; B (\partial_{1} B + e \bar \psi \gamma_{1} \psi)
- {\cal B} ( \partial_{1} {\cal B} - e \bar \psi \gamma_{0} \psi)\;
\bigr ], \nonumber\\
&\equiv& {\displaystyle \int} dx\; \bigl [\; B \dot {\cal B} -
\dot B {\cal B} \;\bigr ],
\end{array}\eqno(12)
$$
where the latter expression for $W$ has been obtained due to the use
of equation of motion: $ \varepsilon^{\mu\nu} \partial_{\nu}
{\cal B} + \partial^{\mu} B = - e \bar \psi \gamma^{\mu} \psi$.
There are other simpler ways to derive the expression for $W$. For
instance, it can be seen that anticommutator of $Q_{B}$ and $Q_{D}$
leads to the derivation of $W$ (i.e. $ \{Q_{B}, Q_{D} \} = W$) if we
exploit the basic (anti)commutators of equation (9). Furthermore,
since $Q_{B}$ and $Q_{D}$ are generators for the transformations
(3) and (7) respectively, it can be seen that the following 
relationships
$$
\begin{array}{lcl}
\delta_{D} Q_{B} &=& - i \eta \{ Q_{B}, Q_{D} \} = - i \eta W,
\nonumber\\
\delta_{B} Q_{D} &=& - i \eta \{ Q_{D}, Q_{B} \} = - i \eta W,
\end{array}\eqno (13)
$$
lead to the definition and derivation of $W$. Together, all the above
generators obey the following extended BRST algebra
$$
\begin{array}{lcl}
&& [ W, Q_{k} ] = 0, \;\;\;\qquad\;\;\; k = g, B, D, AB, AD, \nonumber\\
&& Q_{B}^2 = Q_{D}^2 = Q_{AB}^2 = Q_{AD}^2 = 0, \qquad
\{ Q_{D}, Q_{AD} \} = 0, \nonumber\\
&& \{ Q_{B}, Q_{D} \} = \{Q_{AB}, Q_{AD} \} = W, \qquad
\{ Q_{B}, Q_{AB} \} = 0, \nonumber\\
&& i [ Q_{g}, Q_{B} ] = Q_{B}, \qquad i [ Q_{g}, Q_{AB} ] = - Q_{AB},
\nonumber\\
&& i [ Q_{g}, Q_{D} ] = - Q_{D}, \qquad i [ Q_{g}, Q_{AD} ] = Q_{AD},
\end{array}\eqno(14)
$$
and the rest of the (anti)commutators are zero. Here $Q_{AD}$ is the
anti-dual BRST charge which can be readily derived from (8) by the
replacement :$ \bar C \rightarrow \pm i C$. It is evident that the
operator $W$ is the Casimir operator for the whole algebra. The 
{\it mathematical aspects} of the representation theory for the above kind of 
BRST algebra
have been discussed in Refs. [16-18]. It will be noticed that the ghost
number for $Q_{B}$ and $Q_{AD}$ is $+1$ and that of $Q_{D}$ and $Q_{AB}$
is $-1$. Now, given a state $| \phi> $  with ghost number $n$
 in the quantum Hilbert space ( i.e.  $ i Q_{g} | \phi > = n |\phi> $),
it can be readily seen, using the algebra (14), that
$$
\begin{array}{lcl}
i Q_{g} Q_{B} | \phi > &=& (n + 1) Q_{B} |\phi>, 
\qquad i Q_{g} Q_{AD} | \phi > = (n + 1) Q_{AD} | \phi >,\nonumber\\
i Q_{g} Q_{D} | \phi > &=& (n - 1) Q_{D} |\phi>,
\qquad i Q_{g} Q_{AB} | \phi > = (n - 1) Q_{AB} | \phi >,\nonumber\\
i Q_{g} W   \;| \phi > &=& n \;    W \;  |\phi>.
\end{array}\eqno(15)
$$   
This shows that the ghost numbers for the 
states $Q_{B} |\phi>$ (or $ Q_{AD} | \phi >$), $Q_{D} |\phi>$ (or
$ Q_{AB} | \phi > $) and $W |\phi>$ in the quantum Hilbert space 
are $ (n + 1), (n - 1)$ and $ n $ respectively.

At this stage, it is worth mentioning that in Refs. [19-21], the analogous
expressions for $Q_{B}, Q_{D}, W$ (cf. eqns.(3),(8),(12)) have been
derived for the {\it free} 2D Abelian- and non-Abelian gauge theories
(having no interaction with matter fields). Recently, these results
have also been shown to exist for the {\it free} Abelian two-form gauge
theory in $ ( 3 + 1 )$ dimensions of space-time [22]. The
topological properties of these {\it free} 2D theories
have been shown to be encoded in the vanishing
of the operator $W$ when equations of motion are exploited. On the
contrary, as it turns out, the operator $W$ is defined
off-shell as well as on-shell for the 2D {\it interacting} BRST invariant
$U(1)$ gauge theory. This is because of the fact that even though
$U(1)$ gauge field is topological (i.e. without any propagating
degrees of freedom), it is coupled to the Dirac fields here
and fermionic degrees of freedom are present in the off-shell as well
as on-shell expression for $W$. Thus, the present theory is an example
of an {\it interacting} topological field theory in 2D.

It is interesting to note that the algebra of
$Q_{B}, Q_{D}$ and $W$ in equation (14) is exactly
identical to the corresponding algebra 
for the exterior derivative ($d, \; d^2 = 0$), dual exterior 
derivative ($\delta = \pm * d *, \; \delta^2 = 0$)
and the Laplacian operator ($ \Delta = (d +\delta)^2 = d \delta + \delta d)$
in the context of discussion of the
de Rham cohomology [14,15]. Furthermore, it can be readily seen that 
the operation of these generators on a state with ghost number $n$
(cf. eqn.(15)) is same as the operation of the above cohomological
operators on a differential form of degree $n$. Thus, it 
is clear that  the BRST
cohomology can be defined comprehensively in terms of the above operators
and Hodge decomposition theorem
can be expressed cogently in the quantum Hilbert space of states where any
arbitrary state $ |\phi>_{n}$ (with ghost number $n$) 
can be written as the sum
of a harmonic state $|\omega>_{n}$ ($ W |\omega>_{n} = 0, Q_{B} |\omega>_{n}
= 0, Q_{D} |\omega>_{n} = 0$), a BRST exact state ($Q_{B} |\theta>_{n-1}$)
and a co-BRST exact state ($Q_{D} |\chi>_{n+1}$). 
Mathematically, this statement can be expressed by the following equation
\footnote{ This equation is the analogue of
the Hodge decomposition theorem which states that, on a
compact manifold, any arbitrary $n$-form $f_{n}
( n = 0, 1, 2.....)$ can be written as the sum
of a harmonic form $\omega_{n}$ ($ \Delta \omega_{n} = 0, d \omega_{n} = 0,
\delta \omega_{n} = 0$), an exact form $d g_{n-1}$ and a co-exact form
$\delta h_{n+1}$ as:
$f_{n} = \omega_{n} + d g_{n-1} + \delta h_{n+1}$.}
$$
\begin{array}{lcl}
|\phi>_{n} = |\omega>_{n} +\; Q_{B} |\theta>_{n-1} + \;Q_{D} |\chi>_{n+1}.
\end{array}\eqno(16)
$$
It is obvious, therefore, that the above symmetry generators $Q_{B}, Q_{D}$ 
and $ W$ have their counterparts in differential geometry as the de Rham 
cohomology operators $ d, \delta$ and $ \Delta$ respectively [14,15] for the 
discussion of cohomological aspects of differential forms. It is a peculiarity
of the BRST formalism that the above cohomological operators can be also 
identified with the generators $ Q_{AD}, Q_{AB}$ and $W 
= \{Q_{AB}, Q_{AD} \}$ respectively. Thus, the mapping is: $ (Q_{B}, Q_{AD})
\Leftrightarrow d, \; (Q_{D}, Q_{AB}) \Leftrightarrow \delta, \; W 
= \{ Q_{B}, Q_{D} \} = \{ Q_{AD}, Q_{AB} \} \Leftrightarrow \Delta$.

It will be very useful to explore the impact of this new symmetry
in the context of symmetries of the Green's functions for QED and derive
the analogue of Ward-Takahashi identities. This new symmetry, being connected 
with the analogue of chiral transformation, is expected to play an
important role in throwing some light on the 2D Adler-Bardeen-Jackiw anomaly  
in the framework of BRST cohomology and Hodge decomposition theorem. 
Furthermore, this symmetry might turn out to provide a key tool
in proving the consistency and unitarity of the anomalous gauge 
theory in 2D (see, e.g., Refs. [23, 24] and references therein).  The 
generalization of this new symmetry to 2D non-Abelian gauge theory (having 
local gauge interaction with matter fields) is another future direction that 
can be pursued. The insights gained in these studies might provide a clue for 
the generalization of this new symmetry to physical four dimensional gauge 
theories. These are some of the issues under investigation and a detailed 
discussion will be reported elsewhere [25].

\end{document}